\documentclass{article}
\usepackage{template/spconf,amsmath,amsfonts,epsfig}
\usepackage[ruled,vlined]{algorithm2e}
\usepackage{hyperref}
\usepackage[nameinlink,capitalize]{cleveref}
\usepackage{xcolor}
\usepackage{multirow}
\usepackage{graphicx}
\usepackage{float}
\usepackage{fancyhdr}
\let\OLDthebibliography\thebibliography
\renewcommand\thebibliography[1]{
  \OLDthebibliography{#1}
  \setlength{\parskip}{0pt}
  \setlength{\itemsep}{0pt plus 0.3ex}
}

\fancypagestyle{FirstPage}{
}

\begin{document}\sloppy

\title{Learned Image Coding for Machines: A Content-Adaptive Approach}
%
\name{\begin{tabular}{c}Nam Le$^{\ast\dag}$, Honglei Zhang$^{\dag}$, Francesco Cricri$^{\dag}$, 
Ramin Ghaznavi-Youvalari$^{\dag}$, \\
Hamed Rezazadegan Tavakoli$^{\dag}$, Esa Rahtu$^{\ast}$\end{tabular}}
\address{$^{\dag}$Nokia Technologies, $^{\ast}$Tampere University \\ 
Tampere, Finland 
}

\maketitle
\newcommand{\fullresincluded}{}

\newcommand{\newparagraph}[1]{\par\textbf{#1}}

\newcommand{\Tensor}[1]{\boldsymbol{#1}}
\newcommand{\Loss}[1]{\mathcal{L}_{#1}}
\newcommand{\LossFT}[1]{\bar{\mathcal{L}}_{#1}}
\newcommand{\Wof}[1]{\bar{w}_{#1}}
\newcommand{\Model}[2]{\boldsymbol{#1}({#2})}
\newcommand{\ModelWeights}[1]{\boldsymbol{\theta}_{\boldsymbol{#1}}}
\newcommand{\Wmodel}[2]{\Model{#1}{{#2};\ModelWeights{#1}}}
\newcommand{\goodresult}[1]{\textcolor{green}{#1}}
\newcommand{\badresult}[1]{\textcolor{red}{#1}}

\renewcommand*{\figureautorefname}{Fig.}
\newcommand{\BDrateWhole}{3.66}
\newcommand{\BDrateLow}{9.85}
\newcommand{\BDrateMid}{0.71}
\newcommand{\BDrateHigh}{0.09}

\newcommand{\BDrateVVCWhole}{25.46}
\newcommand{\BDrateVVCLow}{4.81}
\newcommand{\BDrateVVCMid}{40.90}
\newcommand{\BDrateVVCHigh}{66.42}

\newcommand{\BDrateVVCFullWhole}{30.54}
\newcommand{\BDrateVVCFullLow}{11.64}
\newcommand{\BDrateVVCFullMid}{44.82}
\newcommand{\BDrateVVCFullHigh}{70.51}

\newcommand{\BDrateICASSPWhole}{23.92}
\newcommand{\BDrateICASSPLow}{7.03}
\newcommand{\BDrateICASSPMid}{26.03}
\newcommand{\BDrateICASSPHigh}{44.07}

\newcommand{\litencoder}{E}
\newcommand{\litdecoder}{D}
\newcommand{\litprobmodel}{P}
\newcommand{\litquantizer}{Q}
\newcommand{\litimg}{x}
\newcommand{\litlatent}{y}
\newcommand{\lithyperlatent}{z}
\newcommand{\litprior}{p}
\newcommand{\litweight}{w}
\newcommand{\litweightset}{W}
\newcommand{\litlossrate}{rate}
\newcommand{\litlosstask}{task}
\newcommand{\litlossmse}{mse}
\newcommand{\litlossfinetune}{total}
\newcommand{\litlossperceptual}{proxy}
\newcommand{\img}{\Tensor{\litimg}}
\newcommand{\resimg}{\Tensor{\hat{\litimg}}}
\newcommand{\latent}{\Tensor{\litlatent}}
\newcommand{\iqlatent}{\Tensor{\hat{\litlatent}}}
\newcommand{\tqlatent}{\Tensor{\Tilde{\litlatent}}}
\newcommand{\hyperlatent}{\Tensor{\lithyperlatent}}
\newcommand{\tqhyperlatent}{\Tensor{\Tilde{\lithyperlatent}}}
\newcommand{\iqhyperlatent}{\Tensor{\hat{\lithyperlatent}}}
\newcommand{\prior}[1]{\litprior_{#1}}
\newcommand{\encoder}[1]{\Wmodel{\litencoder}{#1}}
\newcommand{\decoder}[1]{\Wmodel{\litdecoder}{#1}}
\newcommand{\probmodel}[1]{\Wmodel{\litquantizer}{#1}}
\newcommand{\wrate}{\litweight_{\litlossrate}}
\newcommand{\wtask}{\litweight_{\litlosstask}}
\newcommand{\wmse}{\litweight_{\litlossmse}}
\newcommand{\wset}{\mathcal{\litweightset}}
\newcommand{\lossrate}{\Loss{\litlossrate}}
\newcommand{\losstask}{\Loss{\litlosstask}}
\newcommand{\lossmse}{\Loss{\litlossmse}}
\newcommand{\lossfinetune}{\LossFT{\litlossfinetune}}
\newcommand{\lossperceptual}{\LossFT{\litlossperceptual}}
\newcommand{\lossratefinetune}{\LossFT{\litlossrate}}
\newcommand{\wofrate}{\Wof{\litlossrate}}
\newcommand{\wofperceptual}{\Wof{\litlossperceptual}}
\newcommand{\latentfinetune}{\bar{\Tensor{\litlatent}}}

\begin{abstract}
Today, according to the Cisco Annual Internet Report (2018-2023),
the fastest-growing category of Internet traffic is machine-to-machine communication. 
In particular, machine-to-machine communication of images and videos represents a new 
challenge and opens up new perspectives in the context of data 
compression. One possible solution approach consists of adapting current human-targeted image and video coding standards to the use case of machine consumption. Another approach consists of developing completely new 
compression paradigms and architectures for machine-to-machine communications. 
In this paper, we focus on image compression and present an inference-time content-adaptive 
finetuning scheme that optimizes the latent representation of an end-to-end learned image codec, 
aimed at improving the compression efficiency for 
machine-consumption. The conducted experiments targeting instance segmentation task network show that our online finetuning brings an average 
bitrate saving (BD-rate) of -\BDrateWhole \% with respect to our pretrained image codec. In particular, 
at low bitrate points, our proposed method results in a significant bitrate saving of 
-\BDrateLow \%. Overall, our pretrained-and-then-finetuned system achieves -\BDrateVVCFullWhole\% 
BD-rate over the state-of-the-art image/video codec Versatile Video Coding (VVC) on instance segmentation.
\end{abstract}
\thispagestyle{FirstPage}
\section{Introduction}
\label{sec:intro}
It is predicted that half of global connected devices and connections will be for 
machines-to-machine (M2M) communications by 2023 \cite{cisco_annualinternet}.
There has been a tremendous improvement of coding efficiency in the recent iterations 
of the human-oriented video coding standards such as Versatile Video Coding (VVC) \cite{vvc}. 
The performance of such codecs, however, remains questionable in the use cases where 
non-human agents, hereinafter \textit{machines}, are the first-class consumers. 
To understand this problem, the Moving Picture Experts Group (MPEG) has recently issued a 
new Ad-hoc group called Video coding for machines (VCM) \cite{vcm_requirements} in the effort to study and standardize 
these use cases. 

In response to this emerging challenge, many studies have been actively conducted to explore 
alternative coding solutions for the new use cases. There exist mainly two categories of 
solutions: adapting the traditional image and video codecs for machine-consumption \cite
{adapting_JPEGXS,vcm_feature_based}, and employing end-to-end 
learned codecs that are optimized directly for \textit{machines} by taking advantage of the 
neural network (NN) based solutions \cite{icassp_paper, intelICIPSemPreserving2020}. Each 
approach has its own pros and 
cons. Traditional video codec-based solutions, built upon mature technologies and broadly 
adopted standards, are often compatible with existing systems \cite{adapting_JPEGXS,vcm_feature_based}. 
However, it is difficult to optimize the overall performance of a system that consists of a 
traditional video codec and neural networks that perform machine tasks \cite{icassp_paper}. 
On the other hand, NN-based codecs are easier to optimize, though 
they may suffer from data domain shift at inference time, when the test data distribution 
differs from the distribution of the data used for training the learnable parameters.



In this work, we address the problem of domain shift at inference time for end-to-end learned 
NN-based solutions. We propose a content-adaptive fine-tuning technique that can enhance the coding 
efficiency on-the-fly at the inference stage without modifying the architecture or the learned parameters 
of the codec. We demonstrate that 
the proposed method, as a complementary process, can improve the performance of a learned image codec 
targeting instance segmentation by saving up to -\BDrateLow\% in bitrate. 

\section{Related work}
\label{sec:related-work}

End-to-end learned approaches aimed at improving the task performance on the outputs of the codec,
have been studied in some recent works. Learned codecs that are optimized for both human and 
machine consumption are proposed in \cite{deepsic} and \cite{intelICIPSemPreserving2020}, 
either by integrating the semantic features into the bitstream \cite{deepsic} or by 
adapting the targeted task networks to the learned latent representation \cite{intelICIPSemPreserving2020}. 
Concentrating solely on machine-consumption use cases, 
the codecs presented in \cite{icassp_paper} achieve state-of-the-art coding performance 
without any modifications to the task networks. The above methods improve the task 
performance by designing new codecs trained on a certain amount of training data, which 
may have a different distribution than that of the test data at inference time. Our proposed 
method adapts only the transmitted data to the input content, hence alleviating
the data domain shift while the codec remains unchanged.

Inference-time finetuning schemes have already been explored in numerous works in the context of 
image compression. The goal is to optimize some parts of the codec in order to maximize its performance 
on the given input test data. This operation is performed on the encoder side, and the subject of 
optimization may be the encoder itself, the output of the encoder, or the decoder. In 
\cite{weightupdate_acm,weightupdate_cvpr}, 
the post-processing filter is finetuned at decoder side by using weight-updates signaled from the encoder 
to the decoder at inference time. Techniques of latent tensor overfitting for better human consumption 
are presented in \cite{jpegai_paper,camposContentAdaptiveOptimization2019a}, aiming at reducing the 
distortion in the pixel domain. Such approaches are useful for enhancing the fidelity 
in the pixel domain and their application is straightforward, as the ground-truth at the encoder-side is 
readily available (the uncompressed image itself). However, similar techniques are not 
directly applicable to a NN-based codec that targets machine tasks, where neither the 
ground-truth nor the task networks are available at the encoder side. In this paper, we 
propose a content-adaptive finetuning technique that can further improve the performance of 
a NN-based codec for machine consumption.

\section{Proposed method}
\label{sec:method}
An image coding for machines (ICM) codec aims to achieve optimal task network performance 
given a targeted bitrate, as opposed to the traditional codecs which are optimized for the 
visual quality. In this work, we propose a content-adaptive finetuning scheme that further 
enhances the coding efficiency of NN-based ICM codecs. As a concrete example, we apply this technique on a baseline ICM codec described below.
\subsection{Baseline ICM system}
\label{ssec:base-codec}
\begin{figure}[t]
    \begin{minipage}[b]{1.0\linewidth}
      \centering
      \centerline{\includegraphics[width=\linewidth]{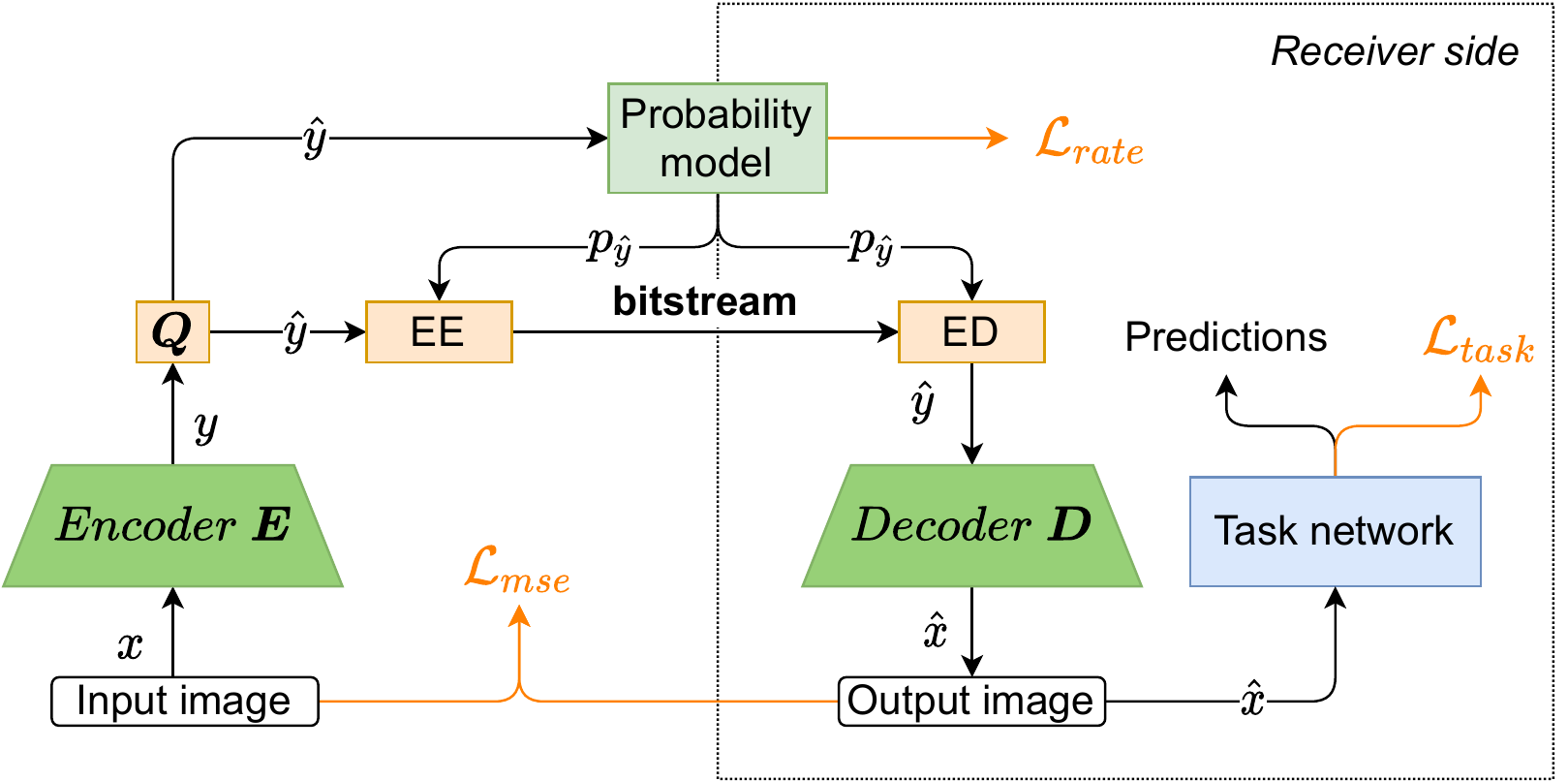}}
    \end{minipage}
  \caption{System architecture of the baseline Image Coding for Machines system.
  ``EE'' and ``ED'' denote arithmetic entropy encoder and decoder, respectively.}
  \label{fig:overview}
\end{figure}

We use an ICM system based on \cite{icassp_paper} as a baseline. Such system consists of 
three main components: a probability model, a task 
network representing the ``machine'', and an autoencoder. The overview of this system is 
presented in \autoref{fig:overview}. To simplify the experiments, while 
still following the same principles proposed in \cite{icassp_paper}, our system has some 
architectural adjustments which are described in the next paragraphs.

\newparagraph{Probability model} estimates the probability distribution and the bitrate of the 
data to be compressed. Instead of the probability model in \cite{icassp_paper}, we incorporate a 
probability model based on the method proposed for a 
lossless image coding system \cite{accv_paper}, in which a $N$-level-downsampled version of the 
input image $z^{(1)}$ is encoded to achieve a higher compression rate. For each level of 
downsampling, the probability model learns the conditional distributions $p(z^{(l)}|z^{(l+1)})$, 
where $z^{(l+1)}$ denotes the downsampled image by a factor of 2 of $z^{(l)}$, and $l \in [1,N]$. 
The bitstream at each level is coded using the corresponding predicted conditional distribution with 
the lower resolution image as the context. 
These bitstreams can perfectly reconstruct $z^{(1)}$ from $z^{(N)}$. We apply this coding 
pipeline to losslessly compress the quantized latent $\iqlatent$ in our base ICM system. The 
bitrate of the coded bitstream is estimated by the Shannon cross-entropy:

\begin{equation}
    \lossrate=\mathbb{E}_{\iqlatent \sim m_{\iqlatent}}\left[-\log_{2} \prior{\iqlatent}(\iqlatent)\right],
    \label{eq:lossrate}
\end{equation}
where $m_{\iqlatent}$ denotes the true distribution of input tensor $\iqlatent$ and 
$\prior{\iqlatent}(\iqlatent)$ denotes the predicted distribution of $\iqlatent$.

\newparagraph{Task network} is based on Mask R-CNN \cite{maskRCNN} for instance 
segmentation. This network is pre-trained and kept unmodified in our experiments. It provides 
predictions for task performance evaluation and task loss $\losstask$ during the 
training of the codec. The task loss is defined as the training loss of the Mask R-CNN network:
\begin{equation}
    \losstask = \underbrace{\Loss{cls} + \Loss{reg}}_{\text{classifier branch}} + \underbrace
    {\Loss{preg} + \Loss{obj}}_{ \text{RPN branch}} + \underbrace{\Loss{mask}}_{\text{mask 
    branch}},
    \label{eq:losstask}
\end{equation}
where $\Loss{cls}, \Loss{reg}, \Loss{preg}, \Loss{obj}$ and $\Loss{mask}$ denote 
classification loss, regression loss, region proposal regression loss, objectness loss and 
mask prediction loss, respectively. These losses come from the 3 branches of the network 
architecture: classifier branch, region proposal network (RPN) branch, and mask predictor branch, respectively. Readers are 
referred to \cite{maskRCNN} for further information about these components.

\begin{figure}[t]
    \begin{minipage}[b]{\linewidth}
      \centering
      \centerline{\includegraphics[trim=0cm 0cm 1.0cm 0cm, 
      width=\linewidth]{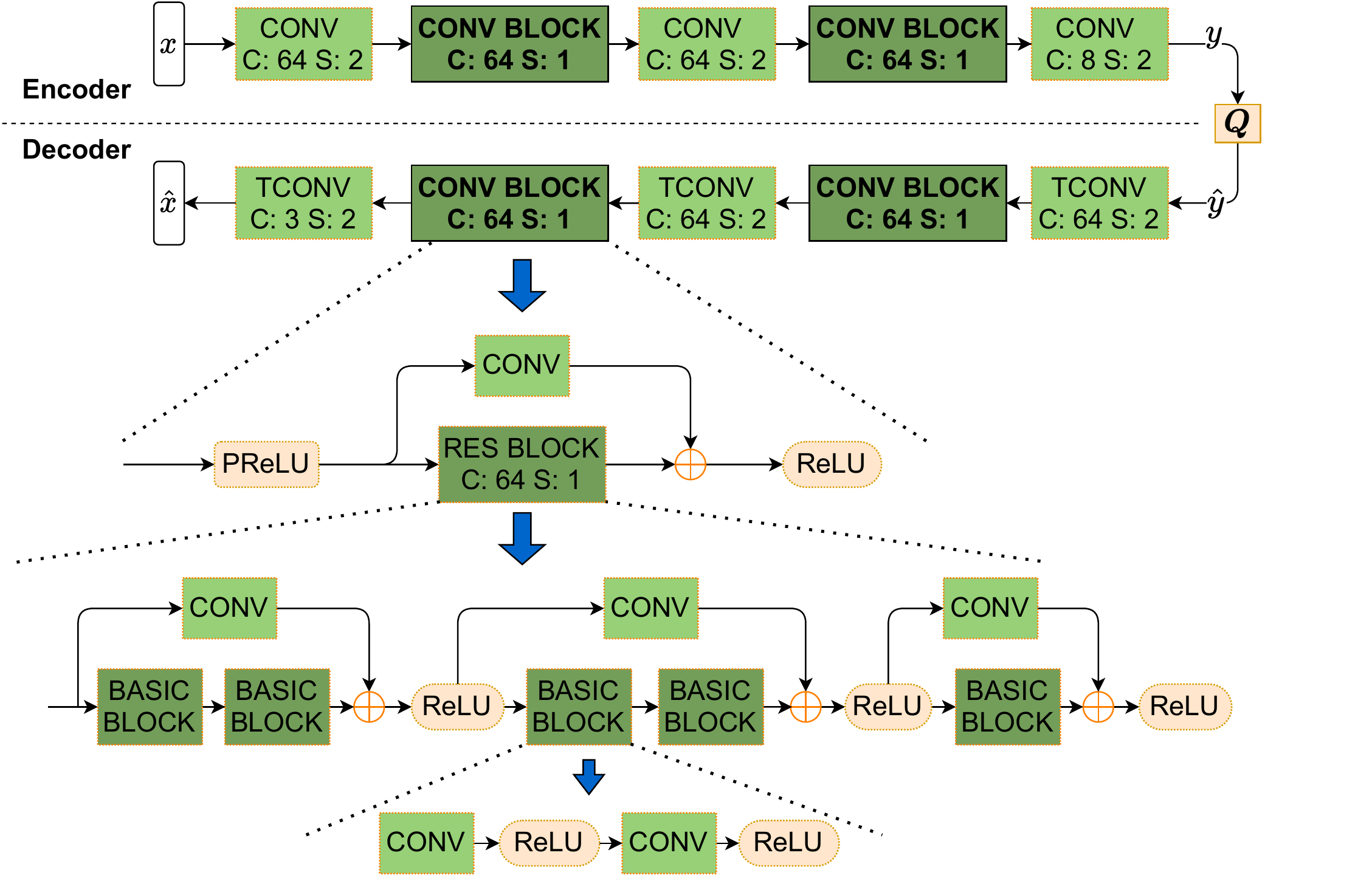}}
    \end{minipage}
\caption{Auto-encoder architecture. The convolutional blocks are illustrated by 
sharp rectangles. ``TCONV'' denotes the transposed convolutional layers. In each 
convolutional block, ``S'' denotes the stride and ``C'' denotes the number of output 
channels for all of the children blocks. These values are inherited from the 
parent block if not stated otherwise.}
\label{fig:autoencoder}
\end{figure}

\newparagraph{Autoencoder} performs the lossy encoding and decoding operations. Together with the probability model and the arithmetic codec, it forms the ``codec'' in the ICM system. It is a 
convolutional neural network (CNN) with residual connections as illustrated in \autoref
{fig:autoencoder}. This is an almost identical network to the one described in \cite
{icassp_paper}, except for the number of channels in the last layer of the encoder. We use a 
6-bit uniform quantizer in this work, denoted as $\Model{Q}{\cdot}$.
The learned encoder takes the uncompressed image $\img$ as input and transforms it into a more 
compressible latent representation $\latent=\encoder{\img}$, where $\ModelWeights{\litencoder}
$ denotes the learned parameters of the encoder. The quantized latent $\iqlatent=\Model{Q}
{\latent}$ is submitted to the probability model for a distribution estimation, which is used 
as the prior for the entropy coding. The decoder transforms the latent data back to the pixel 
domain as the reconstructed image $\resimg=\decoder{\iqlatent}$. 

\newparagraph{Training strategy:} the above ``codec'' is trained to optimize the multi-task 
loss function:
\begin{equation}
    \Loss{train} = \wrate \cdot \lossrate + \wtask \cdot \losstask + \wmse \cdot \lossmse,
\end{equation}
where $\lossrate$ and $\losstask$ are specified by \cref{eq:lossrate} and \cref{eq:losstask}, $\lossmse$ denotes the mean square error 
between $\img$ and $\resimg$, and $\wrate$, $\wtask$, $\wmse$ are the scalar weights for the above 
losses, respectively. These values follow the same configuration proposed in \cite{icassp_paper}, in order to delicately handle the loss terms balancing problem and at the same time train the codec to achieve good efficiency on a wide variety of targeted bitrates.
\subsection{Online latent tensor finetuning}
\begin{figure}[t]
    \begin{minipage}[b]{0.9\linewidth}
        \centering
        \centerline{\hspace*{0.75cm}\includegraphics[width=\linewidth]{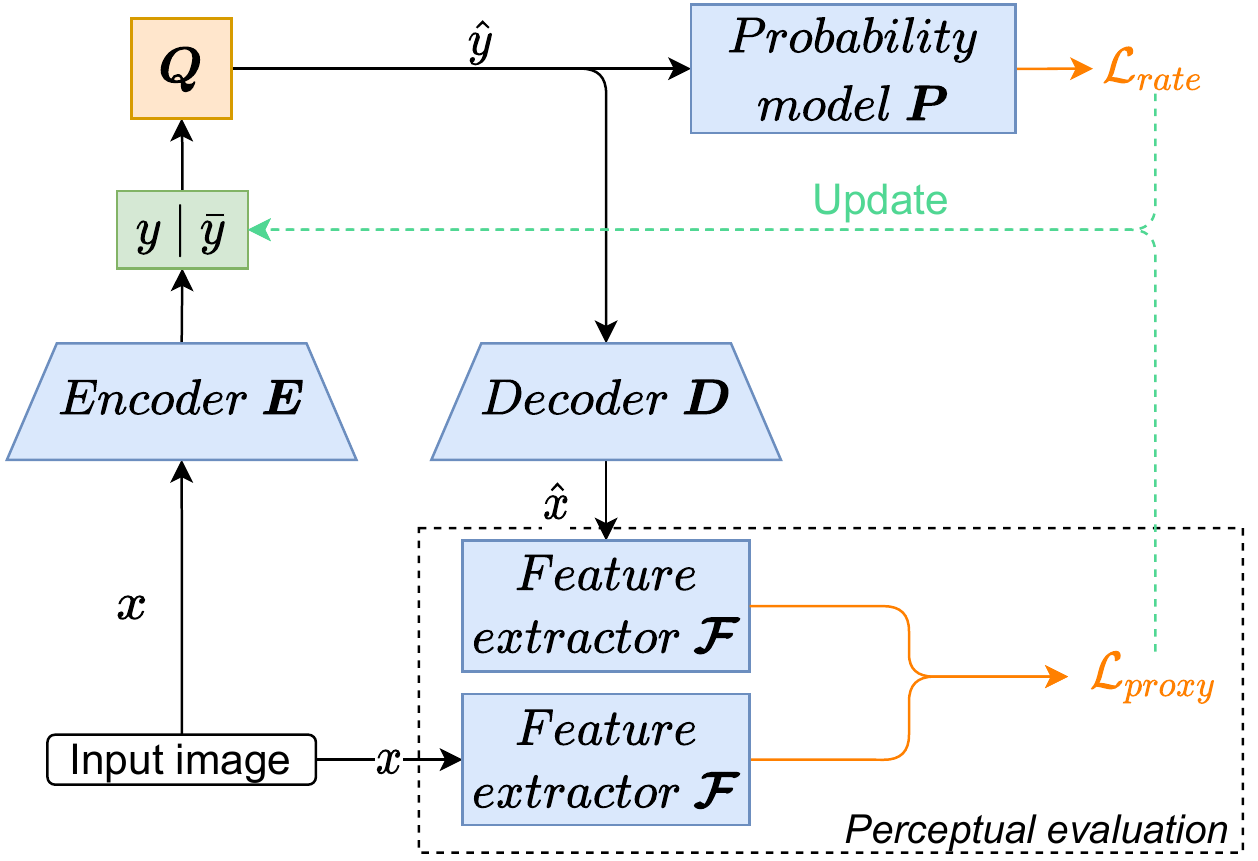}}
    \end{minipage}
    \caption{Online latent tensor finetuning pipeline}
    \label{fig:finetuning-overview}
\end{figure}

\begin{figure}[t]
  \begin{minipage}[b]{\linewidth}
      \centering
      \centerline{\includegraphics[width=\linewidth]{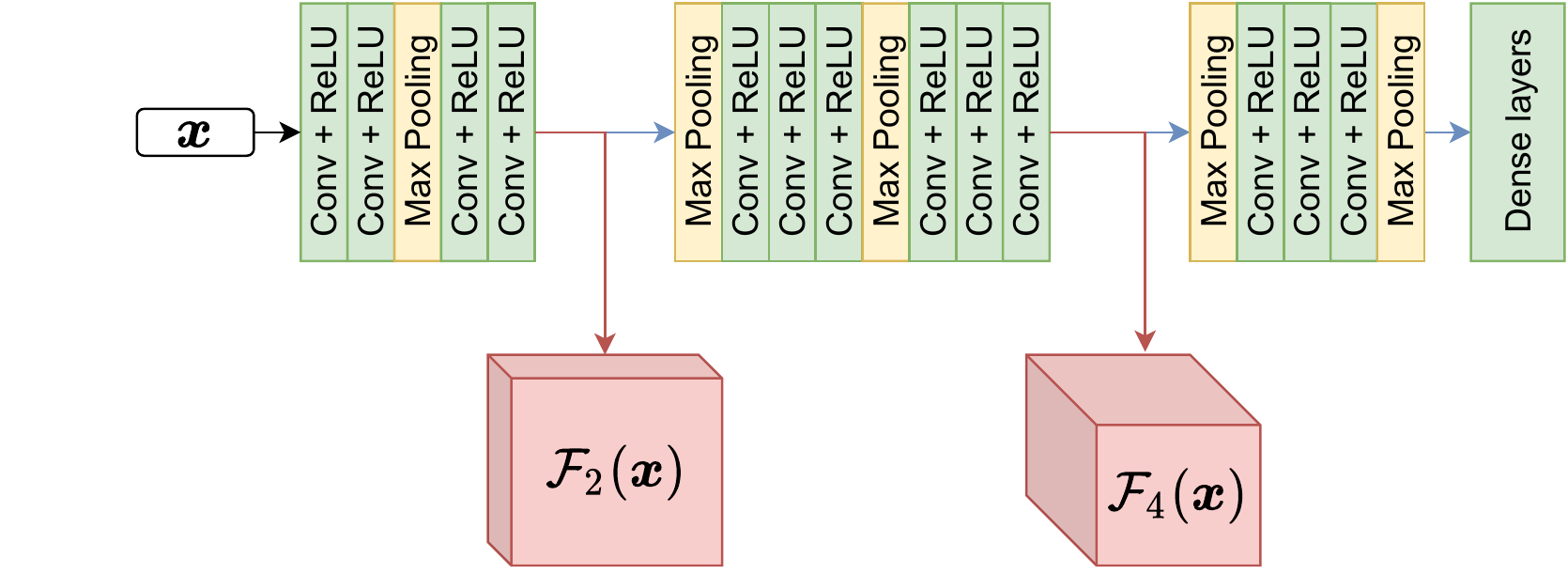}}
  \end{minipage}
  \caption{Feature extraction using VGG-16 (\cref{eq:perceptual-loss}).}
  \label{fig:vgg-extraction}
\end{figure}

At the inference stage, it is possible 
to further optimize the system by adapting it to the content being encoded. 
This way, even better rate-task performance trade-offs can be achieved. 
However, this content-adaptive optimization should be done only for the 
components at the encoder-side, otherwise additional signals containing the updates need to be sent to 
the decoder-side, resulting in bitrate overhead. 
Instead of finetuning the encoder, it is sufficient to finetune 
the latent tensor $\latent$, by back-propagating gradients of the loss with 
respect to elements of $\latent$ through the frozen decoder and probability 
model. 
However, at the inference stage, $\losstask$ is unknown to the encoder since 
the ground-truth for the tasks is not available. Furthermore, the task network is only available on the decoder side and not on the encoder side.  
We hypothesize that the intermediate layer features of the task networks should be 
correlated to those of other vision tasks to a certain degree. 
Therefore, we propose finetuning $\latent$ by using the gradients of the following finetuning loss:
\begin{equation}
    \lossfinetune = \wofrate \cdot \lossratefinetune + \wofperceptual \cdot \lossperceptual, 
\end{equation}
where $\wofrate, \wofperceptual$ denote the weights for the finetuning 
loss terms, $\lossratefinetune$ is similar to $\lossrate$, and $\lossperceptual$ denotes a feature-based perceptual loss which acts as a 
proxy for the task loss $\losstask$. \autoref{fig:finetuning-overview} illustrates our 
finetuning scheme.
In our experiments, we used a VGG-16 \cite{vgg} model pretrained on ImageNet \cite{imagenet}
as the feature extractor. The perceptual loss term $\lossperceptual$ is given by 
\begin{equation}
    \lossperceptual = \mathbf{MSE}(\mathcal{F}_2(\img), \mathcal{F}_2(\resimg))
    + \mathbf{MSE}(\mathcal{F}_4(\img), \mathcal{F}_4(\resimg)),
    \label{eq:perceptual-loss}
\end{equation}
where $\mathbf{MSE}$ denotes the mean square error calculation and $\mathcal{F}_i(\Tensor{t})$ 
denotes the input of the 
$i^{th}$ Max Pooling layer of the feature extractor given the input 
$\Tensor{t}$. The feature extraction is visualized in \autoref{fig:vgg-extraction}.
The finetuning process is described by \autoref{alg:finetuning}.
By using the gradients from $\lossfinetune$ w.r.t $\latent$, the system learns to 
update $\latent$ for better coding efficiency without the knowledge of the task network 
or ground-truth annotations, which makes it a sensible solution.

\begin{algorithm}[h]
  \SetKwInput{KwInput}{Input}                
  \SetKwInput{KwOutput}{Output}              
  \DontPrintSemicolon
  \SetAlgoLined
  \KwInput{Input image $\img$, learning rate $\eta$, number of iterations $n$, loss weights $\wofrate$, $\wofperceptual$}
  \KwOutput{Content-adapted latent tensor $\latentfinetune$}
  \SetKwProg{Fn}{function}{:}{}
  
  \SetKwFunction{FFL}{FeatureLoss}
  \SetKwFunction{FMain}{Main}
  \;
  \Fn{\FFL{$\Tensor{t}_1$, $\Tensor{t}_2$}}{

      \KwRet $\mathbf{MSE}(\mathcal{F}_2(\Tensor{t}_1), \mathcal{F}_2(\Tensor{t}_2))
      + \mathbf{MSE}(\mathcal{F}_4(\Tensor{t}_1), \mathcal{F}_4(\Tensor{t}_2))$\;
  }\;
  $\latentfinetune$ = $\Model{\litencoder}{\img}$ \;
  \tcp{Finetuning iterations}
  \For{$i$ = $1 \to n$}
  {
    $\iqlatent$ = $\Model{\litquantizer}{\latentfinetune}$ \;
    $\prior{\iqlatent}$ = $\Model{\litprobmodel}{\iqlatent}$ \;
    $\resimg$ = $\Model{\litdecoder}{\iqlatent}$ \;
    $\lossratefinetune$ = $\mathbb{E}_{\iqlatent \sim m_{\iqlatent}}\left[-\log_{2} \prior{\iqlatent}(\iqlatent)\right]$ \;
    $\lossperceptual$ = \FFL($\img$, $\resimg$) \;
    $\lossfinetune$ = $\wofrate \cdot \lossratefinetune + \wofperceptual \cdot \lossperceptual$ \;
    $\latentfinetune$ = $\latentfinetune - \eta \cdot \nabla_{\latentfinetune}\lossfinetune$
  }
  \KwRet $\latentfinetune$\;
  \caption{Latent finetuning at inference stage}
  \label{alg:finetuning}
\end{algorithm}
  
\section{Experiments}
\label{sec:experiments}
\subsection{Experimental setup}
For the baseline system, we followed the same training setup as in \cite{icassp_paper}. We used 
Cityscapes dataset \cite{cityscapes} which contains uncompressed images of resolution $2048 
\times 1024$, in two subsets: \textit{train} and \textit{val}. We trained 
the baseline system described in the previous section on the \textit{train} set of 2975 images 
for these classes: \textit{car, person, bicycle, bus, truck, train, motorcycle}. The 
instance segmentation task network is provided 
by Torchvision\footnote{The pre-trained models can be found at 
\url{https://pytorch.org/docs/stable/torchvision/models.html}}. The system is implemented with Pytorch 1.5.

\begin{figure}[t]
  \begin{minipage}[b]{\linewidth}
    \centering
    \ifdefined\fullresincluded
      \centerline{\includegraphics[width=\linewidth]{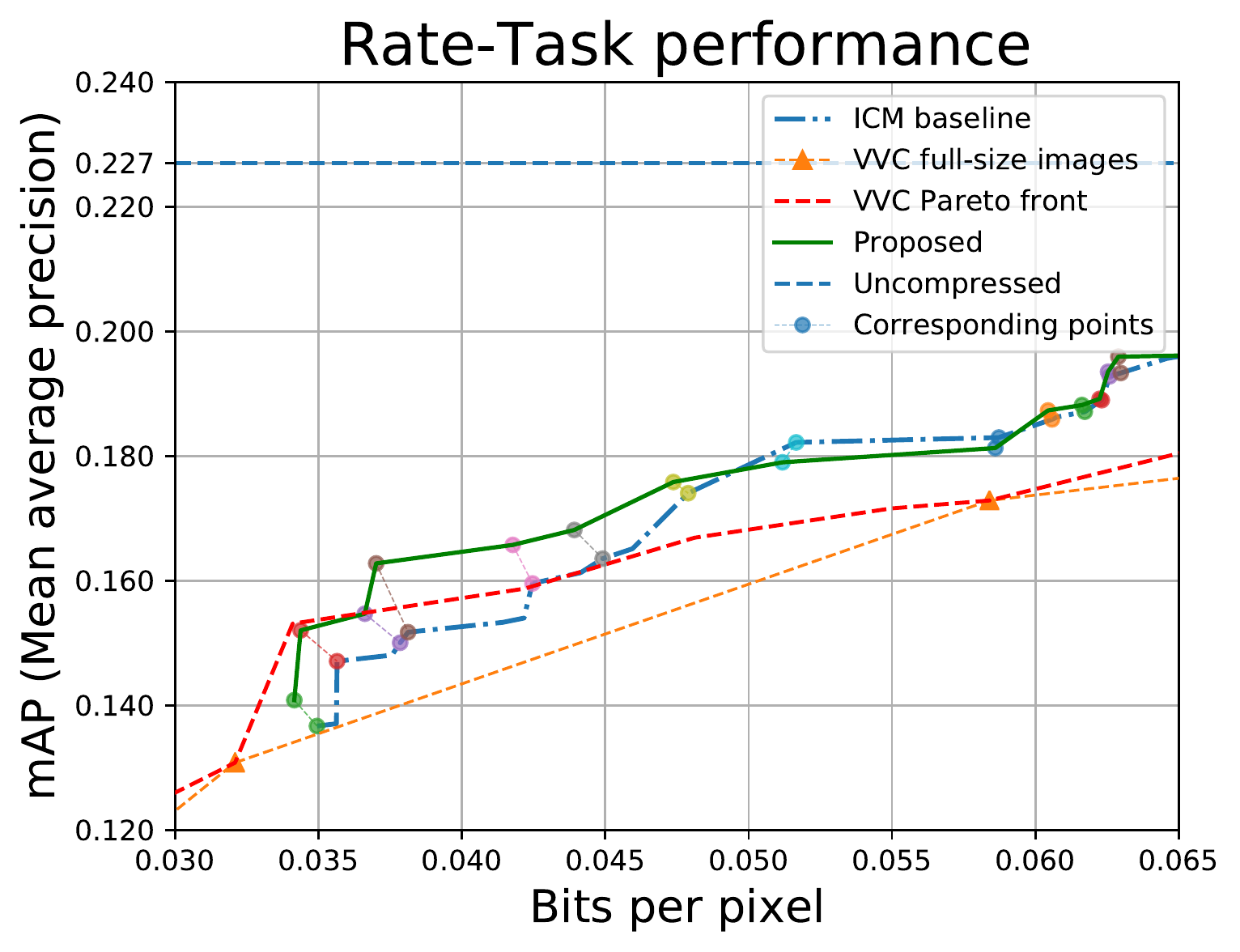}}
    \else
      \centerline{\includegraphics[width=\linewidth]{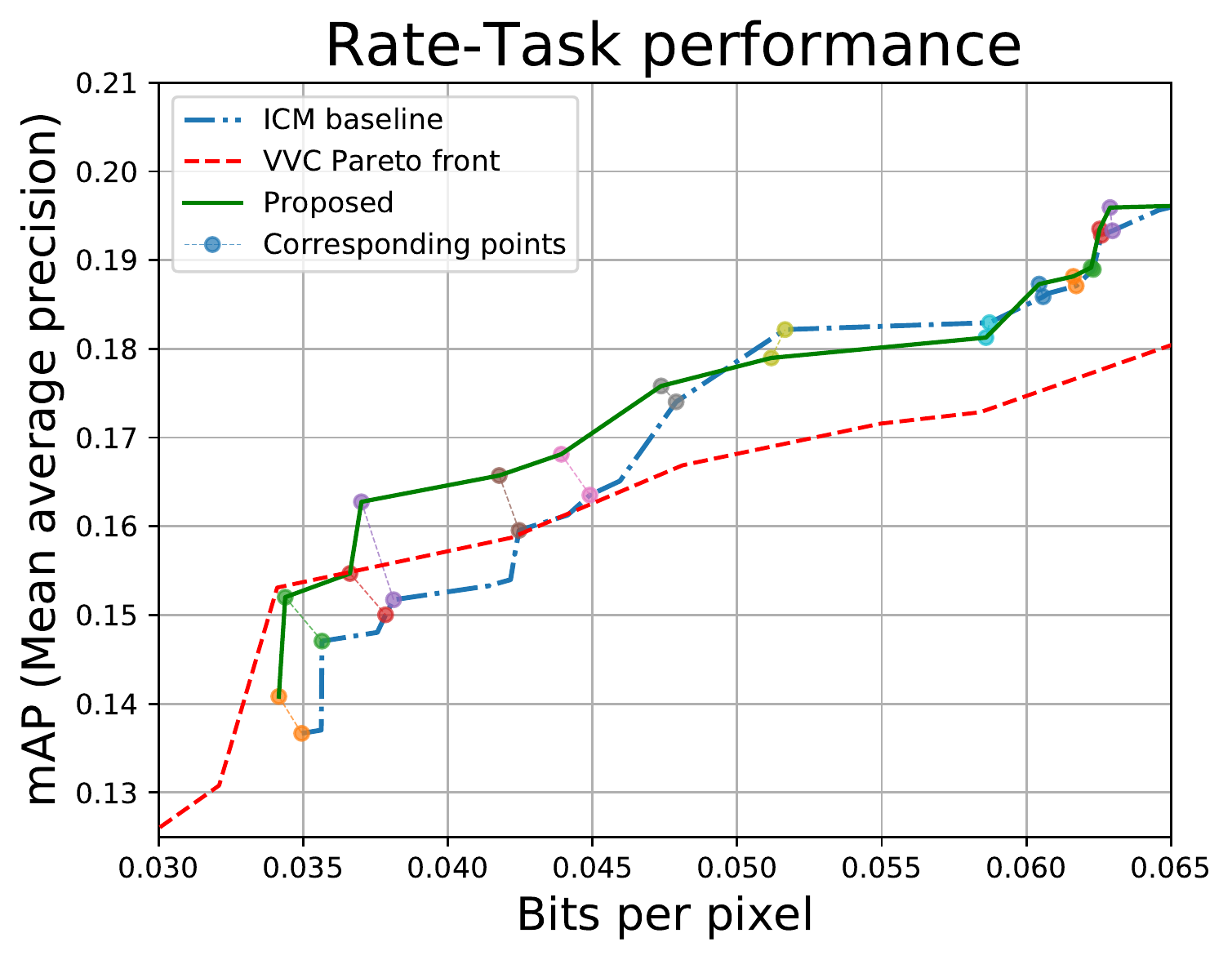}}
    \fi
  \end{minipage}
\caption{Rate-performance curves on low bitrates. The small dashed lines connect the baseline checkpoints and their respective finetuned versions.}
\label{fig:rd-low}
\end{figure}

After every training epoch, we evaluated the coding efficiency with respect to the task 
performance of the model on 500 images of \textit{val} set. 
We use mean average precision (mAP@[0.5:0.05:0.95], described in \cite{cityscapes}) and 
bits per pixel (BPP) as the metrics for task performance and bitrate, respectively.  Note 
that because of the varying weights for the loss terms throughout the training, the codec is 
able to provide different trade-offs that cover a wide range of bitrates. A Pareto front set 
of these checkpoints (i.e., saved models of the epochs) was selected and visualized in 
\autoref{fig:rd-low} as ``ICM baseline''. For simplicity, 
we report the bitrate estimations using the Shannon entropy in \cref{eq:lossrate} instead
of the bitrates arising from the actual bitstream lengths. In practice, we have verified that the differences 
between them are negligible.

We also provide evaluation results using the state-of-the-art video codec VVC
(reference software VTM-8.2\footnote{\url{https://vcgit.hhi.fraunhofer.de/jvetVVCSoftware_VTM}}, All-intra configuration), under JVET common 
test conditions \cite{ctc}. The \textit{val} set is coded in 28 settings of 7 quantization 
parameters (22, 27, 32, 37, 42, 47 and 52) and 4 downsampling factors (1, 0.75, 0.50 and 0.25) 
to achieve different output bitrates. The 28 coded versions of the original input are then 
evaluated for task performance. The Pareto front from these 28 pairs of bitrate and 
corresponding task performance is clipped for the relevant bitrate range and visualized in 
\autoref{fig:rd-low} as ``VVC Pareto front''.
\ifdefined\fullresincluded
Additionally, ``VVC full-size images'' is illustrated representing only the data points that are coded without downsampling, i.e. common use case of VVC.
\fi

Since the baseline model already has a significant performance gain in the higher bitrates, we focus on finetuning 
the coding efficiency at low bitrates in this work. Thus, we used the learning rate $\eta=10^{-4}$ 
to prevent large changes to latent entropy values, which correspond to the bitrates. We observed that the 
losses converge after a few iterations on many samples, therefore the number of iterations $n=30$ and loss 
weights $\wofrate=1$ and $\wofperceptual=0.1$ are empirically chosen in order to achieve a feasible runtime. 
More sophisticated algorithms and strategies can be applied to this process.

At the inference stage, for every baseline checkpoint, we finetune the latent 
$\latent$ of images in the \textit{val} set individually using the Adam 
optimizer \cite{adam}.
Then the finetuned latent $\latentfinetune$ is evaluated. 
The average coding efficiency of the finetuned latents is reported over the whole \textit{val} set for each checkpoint. 
A Pareto front set from the finetuned data points is selected to compare with the 
baseline and is visualized in \autoref{fig:rd-low} as the performance of our method.
\subsection{Experimental results}
The finetuning at inference time offers a bitrate saving, indicated by Bj{\o}ntegaard Delta (BD)-rate 
\cite{bdrate}, of up to -\BDrateLow \%. The average bitrate saving over the whole evaluated 
bitrate 
range is -\BDrateWhole \%. In low bitrates ($<0.05$ BPP), this process significantly boosts the 
coding efficiency as shown in \autoref{fig:rd-low}. 
\begin{table}
  \caption[table caption]{BD-rates w.r.t task performance of our proposed method on different bitrates. 
   }
  \centering
  \resizebox{\linewidth}{!}{
  \setlength\tabcolsep{4pt}
  \begin{tabular}{l|cccc}
  \hline
  \multirow{2}{*}{\textbf{Compared to}} &\multicolumn{4}{c}{\textbf{Bitrate (BPP)}} \\ \cline{2-5}
  & $<0.05$ & $[0.05, 0.1]$ & $ >0.1$ & Total \\ \hline
  ICM baseline& -\BDrateLow\%       & -\BDrateMid\%      & -\BDrateHigh\%                & -\BDrateWhole\%      \\
  VVC Pareto& -\BDrateVVCLow\%       & -\BDrateVVCMid\%      & -\BDrateVVCHigh\%                & -\BDrateVVCWhole\%      \\ 
  \ifdefined\fullresincluded
  VVC full-size&     -\BDrateVVCFullLow\%  &   -\BDrateVVCFullMid\%    &          -\BDrateVVCFullHigh\%       & -\BDrateVVCFullWhole\%      \\ 
  \fi
  \end{tabular}}
  \label{tab:bdrates}
\end{table}

A summary of the BD-rates in different bitrate ranges is presented in \autoref{tab:bdrates}. Apart 
from the coding performance enhancement achieved by the proposed finetuning 
technique, this table also shows impressive BD-rates of up to -\BDrateVVCHigh \% at low bitrates and -\BDrateVVCWhole\% on average when compared to the VVC codec for machine-consumption. As a reference, this measurement of the segmentation codec in \cite{icassp_paper}, which is a different codec than our ICM baseline, measured on the same bitrate range as in our evaluation\footnote{Determined by the minimum and maximum achievable bitrates using our trained models: $[0.034, 0.148]$} is -\BDrateICASSPWhole\% on average. The finetuning of 30 
iterations takes around 2 hours for each checkpoint on the whole $val$ set, or $\approx 14$ 
seconds per image, with a NVIDIA RTX 2080Ti GPU.

\begin{figure*}[ht]
    \begin{minipage}[b]{0.32\linewidth}
      \centering
      \centerline{\includegraphics[width=\linewidth]{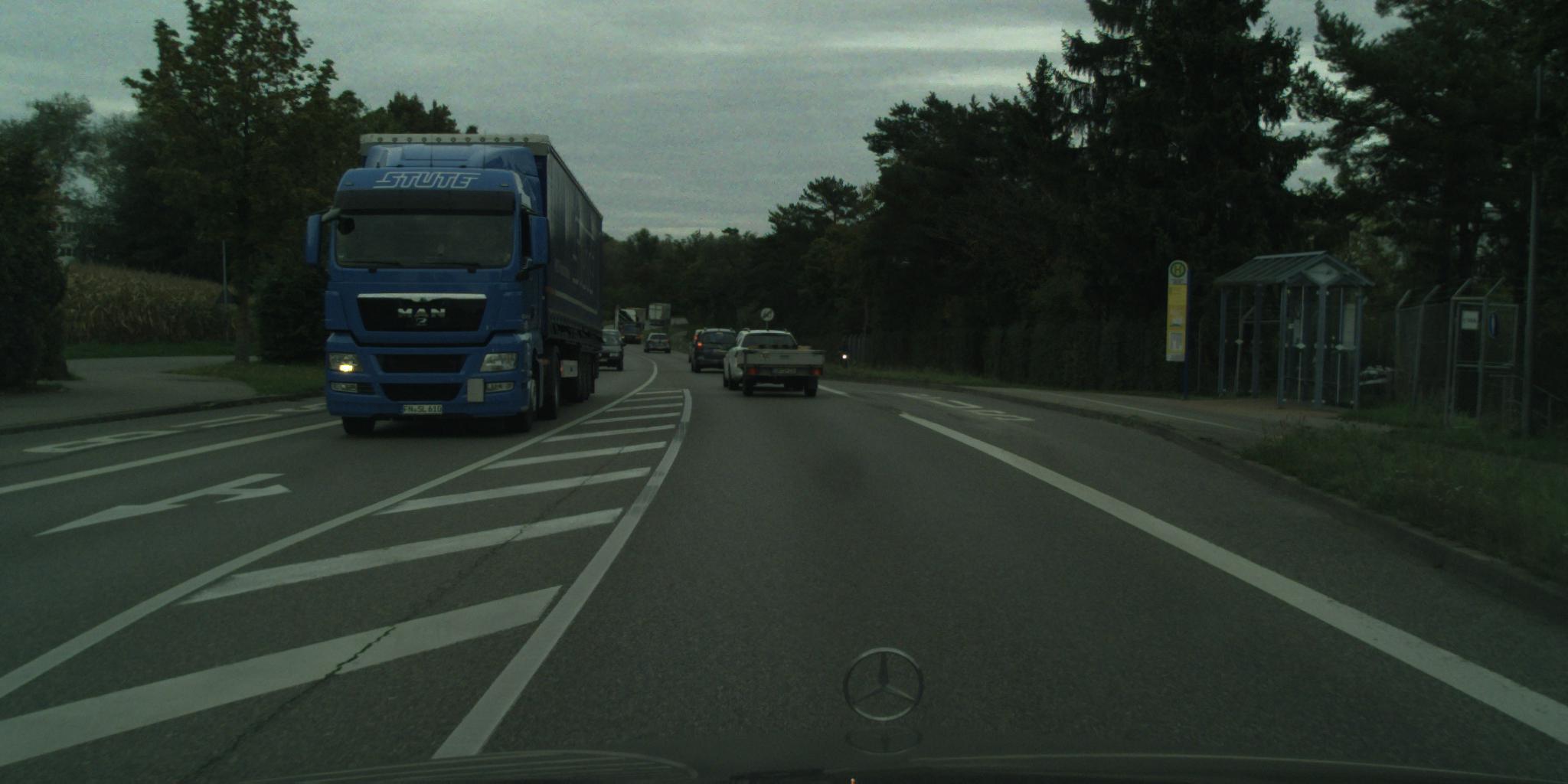}}
    \centerline{Input} 
    \centerline{} 
    \end{minipage}
    \begin{minipage}[b]{0.32\linewidth}
        \centering
        \centerline{\includegraphics[width=\linewidth]{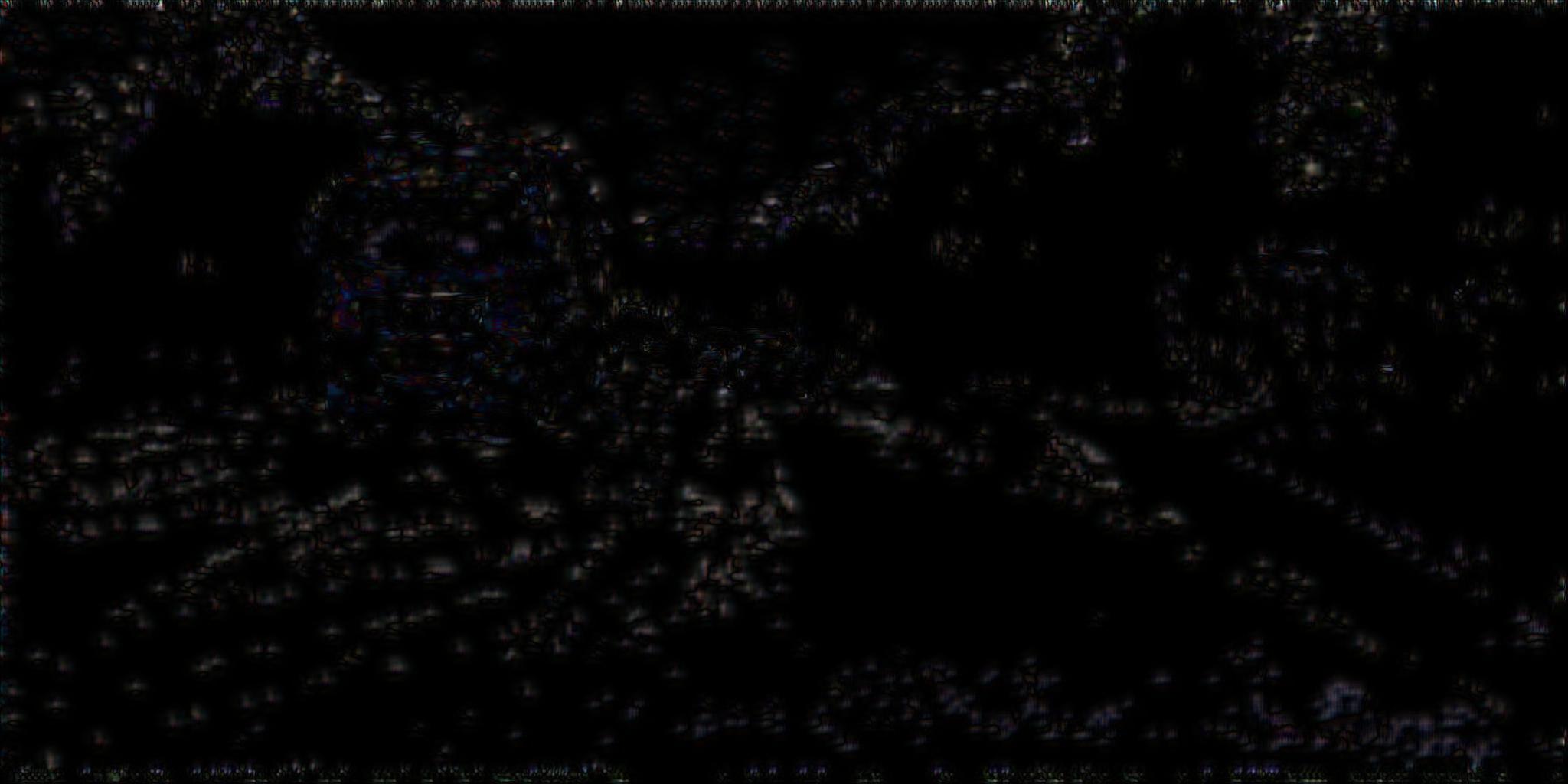}}
      \centerline{Before: 0.026 BPP -- 0.281 mAP}
      \centerline{After: \goodresult{$\downarrow 4.332\%$} -- \goodresult{$\uparrow 6.979\%$}}
    \end{minipage}
    \begin{minipage}[b]{0.32\linewidth}
      \centering
      \centerline{\includegraphics[width=\linewidth]{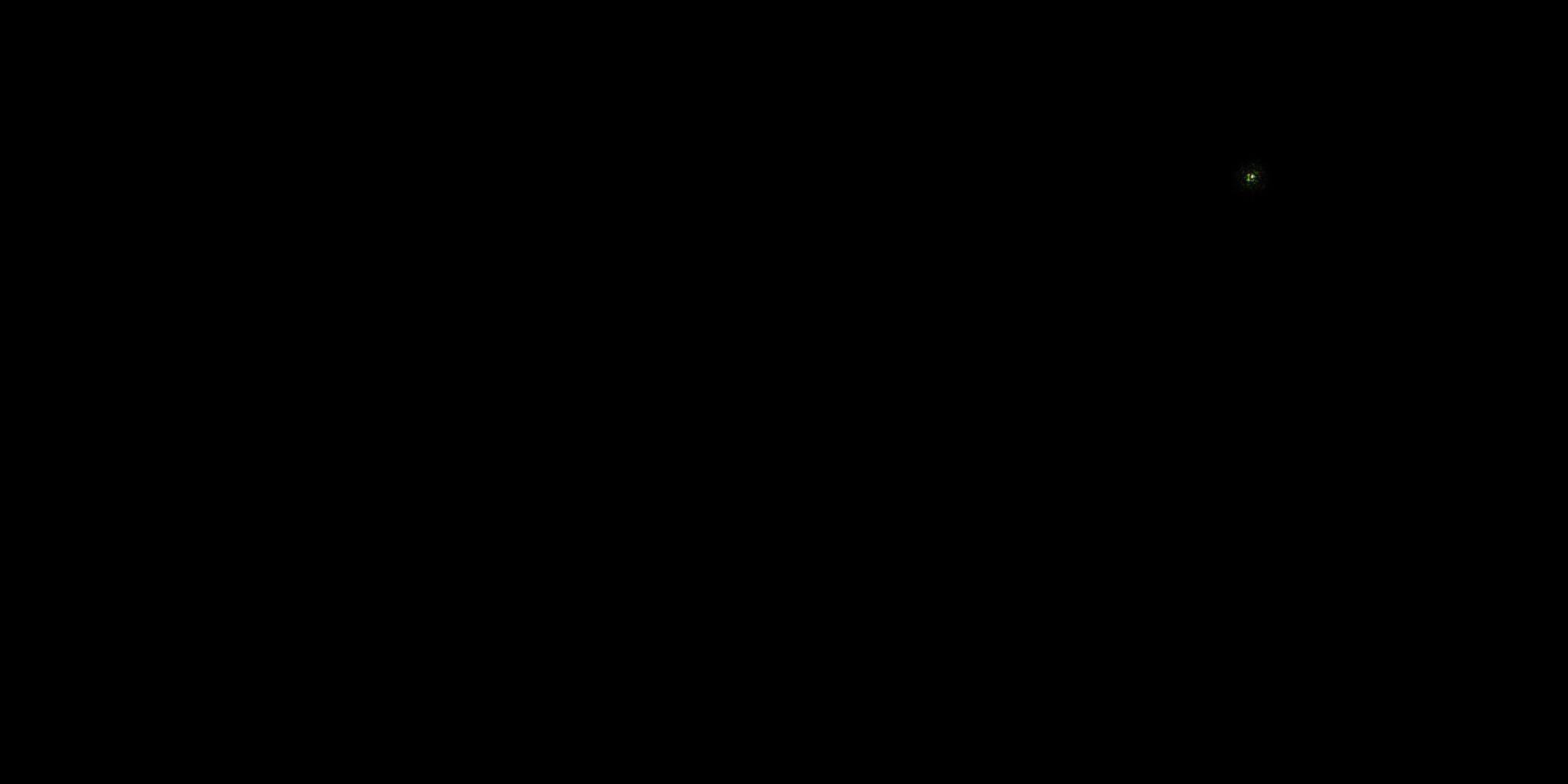}}
      \centerline{Before: 0.131 BPP -- 0.297 mAP}
      \centerline{After: \badresult{$\uparrow 1.6{\times}10^{-4}\%$} -- 0\%}
    \end{minipage}

    \begin{minipage}[b]{0.32\linewidth}
      \centering
      \centerline{\includegraphics[width=\linewidth]{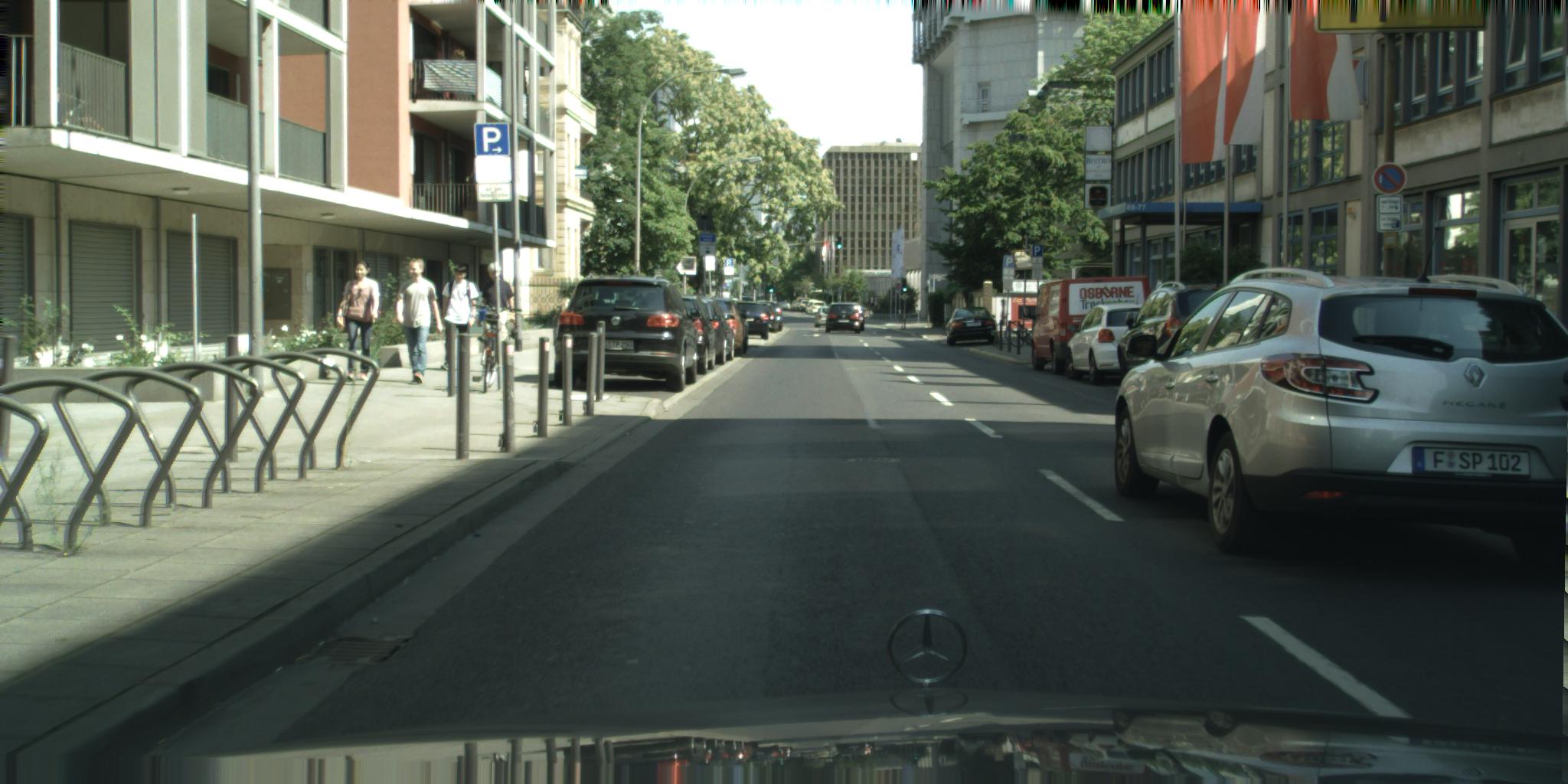}}
    \centerline{Input}
    \centerline{}
    \end{minipage}
    \begin{minipage}[b]{0.32\linewidth}
        \centering
        \centerline{\includegraphics[width=\linewidth]{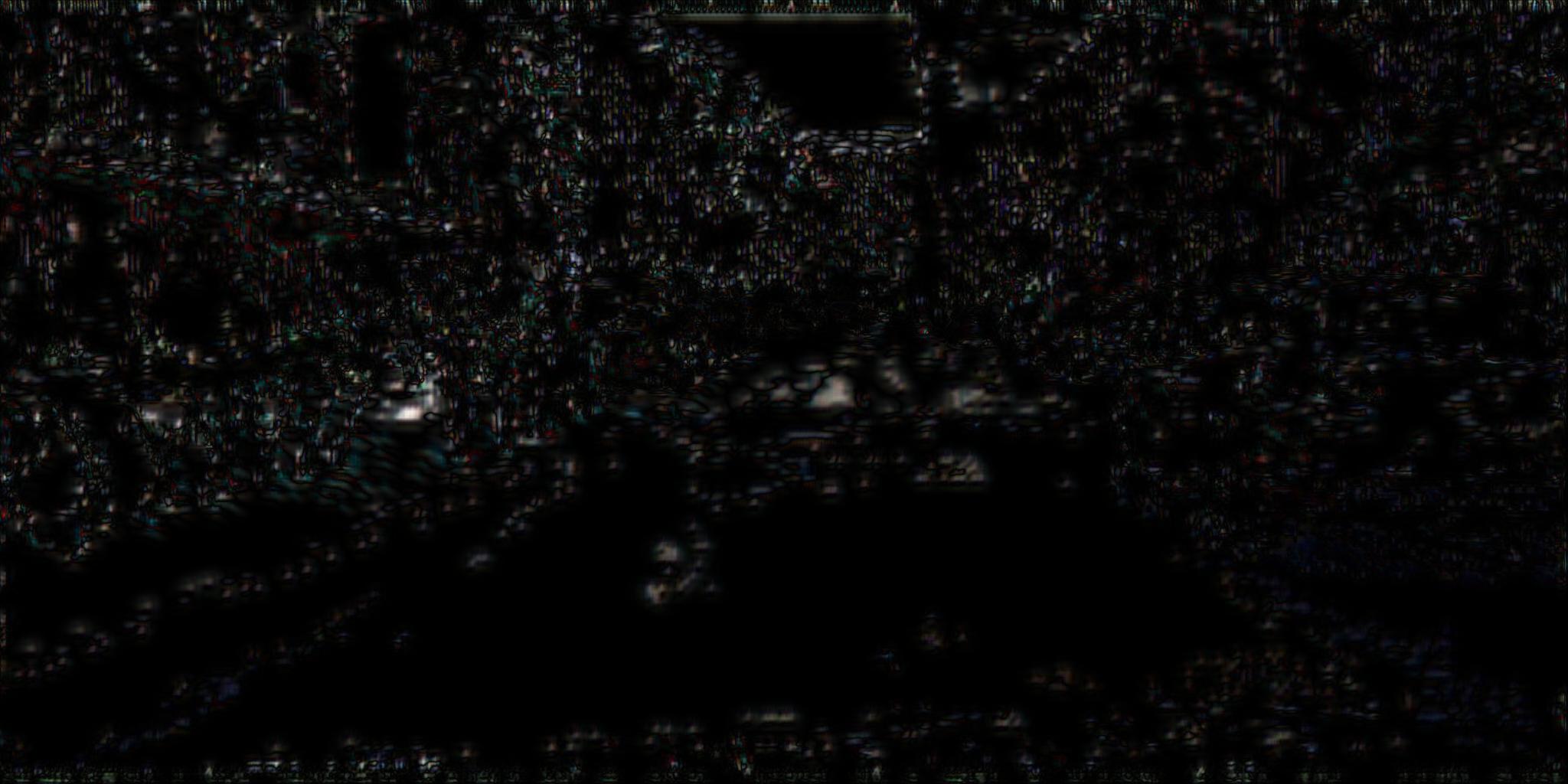}}
        \centerline{Before: 0.043 BPP -- 0.294 mAP}
        \centerline{After: \goodresult{$\downarrow 3.716\%$} -- \goodresult{$\uparrow 2.609\%$}}
    \end{minipage}
    \begin{minipage}[b]{0.32\linewidth}
      \centering
      \centerline{\includegraphics[width=\linewidth]{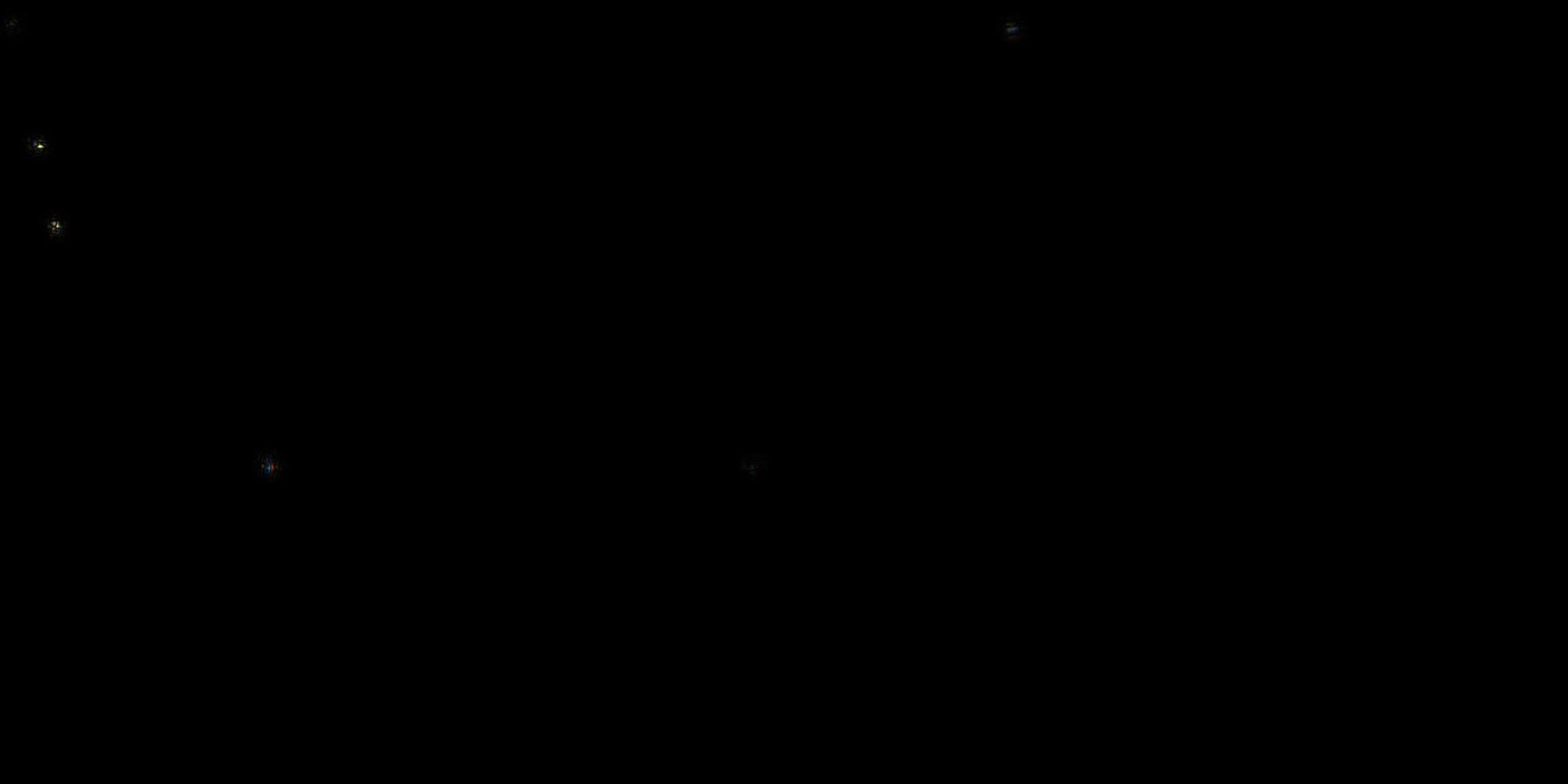}}
    \centerline{Before: 0.123 BPP -- 0.358 mAP}
    \centerline{After: \goodresult{$\downarrow 2.3{\times}10^{-3}\%$} -- 0\%}
    \end{minipage}

    \caption{Finetuning effects on different targeted bitrates. In each row, the leftmost
    image is the uncompressed input and the other images are the $\ell_1$-norm difference 
    images between the baseline outputs and the finetuned ones for different targeted bitrates. The gains after finetuning are given under the images. The finetuning process modifies the areas around the edges 
    and surfaces of the objects in low-bitrate targeted output, which allows for higher coding gains.
    }
    \label{fig:of-outputs}
\end{figure*}

\begin{figure}[ht]
  \centering
  \begin{minipage}[b]{\linewidth}
    \centerline{\includegraphics[width=\linewidth]{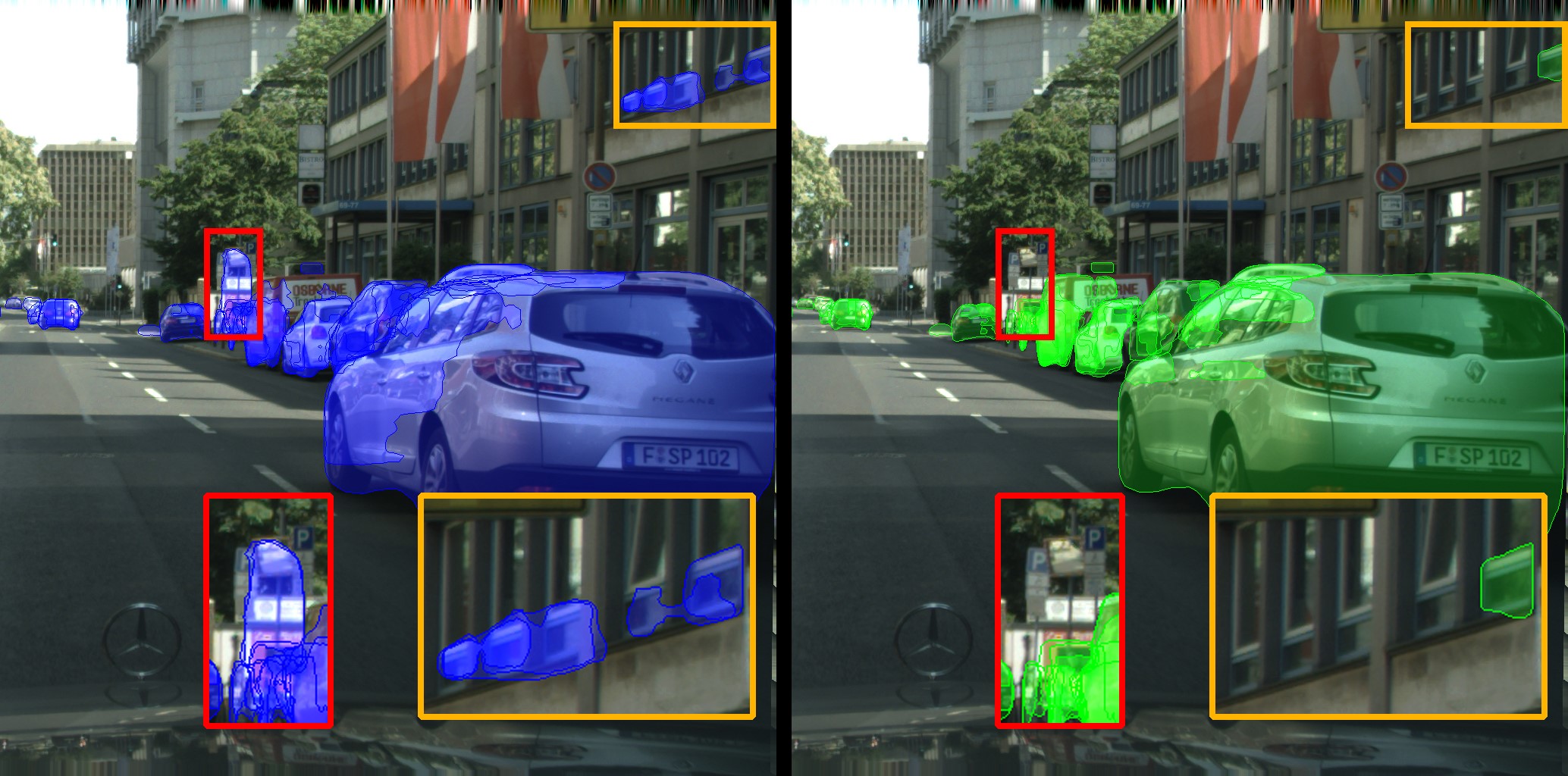}}
  \end{minipage}
\caption{Error reduction of segmentation before and after the finetuning. Left: baseline results, right: finetuned results.}
\label{fig:seg-finetuned}
\end{figure}

In \autoref{fig:of-outputs}, the finetuning on the baseline outputs targeting low bitrates clearly show that 
there are structural modifications around the edge areas of the objects in the images, consequently resulting 
into higher coding performance. 
The outputs targeting high bitrates are only negligibly modified, which aligns 
with the reported insignificant bitrate saving in this range. 
\autoref{fig:seg-finetuned} shows an example of the segmentation error being reduced after finetuning.
\section{Conclusions}
\label{sec:summary}
The proposed content-adaptive finetuning technique can significantly improve the coding 
performance especially for low targeted bitrates. This technique does not require a 
differentiable encoder and it is not dependent on the availability of the task network or task 
ground-truth, therefore it can be easily adopted into most computer vision workflows that employ 
NN-based coding. Additionally, the comparison to the traditional VVC codec confirms the 
superior coding performance of the machine-targeted codec for machine-consumption. 

To further improve our technique, future work could explore 
different configurations to magnify the enhancement of coding efficiency in a 
wider range of bitrates. Furthermore, the effectiveness of this finetuning technique could be 
verified for more computer vision tasks, either separately or on the same finetuned output.

\bibliographystyle{template/IEEEbib}
\bibliography{refs}

\end{document}